\documentclass[12pt]{article} \input epsf.tex

\newcommand{\be}{\begin{equation}}
\newcommand{\ee}{\end{equation}}
\newcommand{\bear}{\begin{eqnarray}}
\newcommand{\eear}{\end{eqnarray}} \newcommand{\ba}{\begin{array}}
\newcommand{\ea}{\end{array}}

\newcommand{\gae}{\begin{array}{c}\,\sim\vspace{-21pt}\\>
\end{array}}

 \newcommand{\CQ}{{\cal Q}}
\newcommand{\CU}{{\cal U}} \newcommand{\CD}{{\cal D}}
\newcommand{\CL}{{\cal L}} \newcommand{\CE}{{\cal E}}
\newcommand{\CN}{{\cal N}}

\begin{document}

\pagestyle{empty} \begin{titlepage}
\def\thepage {}        

\title{\LARGE \bf  Number of Fermion Generations Derived\\ [.3cm]
 from Anomaly Cancellation \\ [.3cm] {} }

\author{\normalsize\bf Bogdan A.~Dobrescu and 
Erich Poppitz\thanks{Address after July 1, 2001: 
Department of Physics, University of Toronto, 
60 St George St., Toronto, ON M5S 1A7 Canada.}
 \\ \\ {\small {\it
\hspace*{-.6cm} Department of Physics, Yale University, New
Haven, CT 06520-8120, USA\thanks{e-mail: bogdan.dobrescu@yale.edu,
erich.poppitz@yale.edu}}}\\
 }

\date{ } \maketitle

   \vspace*{-8cm}
\noindent \makebox[13.2cm][l]{\small \hspace*{-.2cm}
hep-ph/0102010} {\small YCTP-P1-01 } \\
\makebox[11.8cm][l]{\small \hspace*{-.2cm} February 1, 2001 }
\\

 \vspace*{10.5cm}

  \begin{abstract}
{\small We prove that global anomaly cancellation requires
more than one generation of quarks and leptons,
provided that the standard model fields propagate in two
universal extra dimensions.
Furthermore, if the fermions of different generations
have the same gauge charges and chiralities, then
global anomaly cancellation implies that there must be
three generations. }

\end{abstract}

\vfill \end{titlepage}

\baselineskip=18pt \pagestyle{plain} \setcounter{page}{1}


The existence of three generations of quarks and leptons is a major
source of bafflement for particle physics. By contrast, the
particle content within a generation is constrained by the
mathematical structure of the standard model. Local anomaly
cancellation substantially reduces the arbitrariness in choosing
the $SU(3)_C
\times SU(2)_W \times U(1)_Y$ charges of the fermions
\cite{Bouchiat:1972iq}. For example, the existence of the
observed quarks requires the leptons to cancel the $SU(2)_W \times
U(1)_Y$ triangle anomalies. Furthermore, the $SU(2)_W$-doublet
lepton cancels the global anomaly \cite{Witten:1982fp} of the
$SU(2)_W$-doublet quark within each generation.

In this letter, we show that the number of generations may also be
determined by the anomaly cancellation conditions. In order for
this to happen, we are led to consider the existence of extra
spatial dimensions accessible to all the standard model particles.
 If the number $\delta$ of these  ``universal'' extra
dimensions is odd, then there are no local gauge anomalies in the
$(4+\delta)$-dimensional theory.\footnote{For odd $\delta$, the
higher-dimensional analog of the 3-dimensional Redlich anomaly
\cite{Redlich:1984kn} could spoil gauge invariance of the quantum
effective action. However, the anomalous variation of the action
can be cancelled by a Chern-Simons term (see
\cite{Intriligator:1997pq} for a discussion of the $\delta = 1$
case.)}
 Therefore, additional anomaly cancellation conditions that may
restrict the number of generations could arise only for even
$\delta$. The natural choice is then $\delta = 2$. Current
experimental data impose a rather loose upper bound $R
\gae (0.5 \; {\rm TeV})^{-1}$ on the size of two universal extra
dimensions \cite{Appelquist:2000nn}.

The Lorentz group in six dimensions has spinorial representations
of definite chirality with four components. A representation of the
$8\times 8$ anti-commuting gamma matrices, $\Gamma^\alpha$ with
$\alpha = 0,...,5$, is given in ref.~\cite{Arkani-Hamed:2000hv}.
The $\Gamma_7$ matrix, analogue to the $\gamma_5$ matrix in four
dimensions, has eigenvalues $\pm 1$ corresponding to the
6-dimensional fermion chiralities. 
A 6-dimensional chiral
fermion, upon compactification on a smooth manifold (without magnetic
fluxes \cite{Frere:2000dc})
to four dimensions, gives rise to  vector-like fermions.
A 4-dimensional theory with chiral fermions can be obtained
by compactifying the two extra dimensions on an orbifold,
 for example, the $T^2/Z_2$ orbifold
constructed in ref.~\cite{Appelquist:2000nn}.
This orbifold gives
rise to a chiral 4-dimensional theory by projecting out half of
the components of the 6-dimensional Weyl fermions, while the gauge
group in four dimensions is the same as in six dimensions---the
standard model gauge group.

We assume that the 6-dimensional theory is chiral and free of
irreducible local as well as global gauge anomalies. Furthermore,
the reducible anomalies are cancelled by the Green-Schwarz
mechanism \cite{Green:1984sg}, which is a generic feature of 
6-dimensional theories \cite{Sagnotti:1992qw}. Our main results refer 
to the non-supersymmetric standard model in six dimensions, but we also
discuss supersymmetry towards the end,
where we point out that it is hard to cancel the anomalies in this
case.

We emphasize that we
consider the 6-dimensional theory and the orbifold construction
in an effective low-energy field theory framework. It would be 
very interesting to 
find an explicit string theory realization of this non-supersymmetric 
field theory. In this
context, we note the assumption that there are no ``twisted sector"
chiral 4-dimensional fermions localized at the fixed points and
charged under the gauge group. These commonly arise in heterotic orbifolds
\cite{Dixon:1985jw}, but not in open-string orbifolds
\cite{Douglas:1996sw}; this suggests that the place to look for
string realization might be a type-I construction, or, perhaps even
a more exotic construction involving little string theory.
However, this lies beyond the scope of this note---all
we aim here is to provide a consistent low-energy framework.

To this end, consider a generation of 6-dimensional fermions,
$\CQ,
\CU, \CD, \CL, \CE$, whose zero modes form a generation of
4-dimensional quarks, $\CQ^{(0)}\equiv (u, d)_L, \;
\CU^{(0)} = u_R, \; \CD^{(0)} = d_R$, and leptons,
$\CL^{(0)} = (e, \nu_e)_L$, $\CE^{(0)} = e_R$. The 4-dimensional
anomalies cancel automatically within a generation, and from a
4-dimensional point of view this is sufficient for consistency. In
what follows, we will assume that the 4-dimensional theory is
obtained as a deformation of a consistent ({\it i.e.}, anomaly
free) 6-dimensional theory. We will show that the 6-dimensional
anomalies do not cancel so easily, and not only restrict the
6-dimensional chiralities within a generation, but also  impose a
constraint on the number of generations.

The local gauge anomaly in six
dimensions is given by a square one-loop diagram (for a
self-contained introduction to anomalies in six dimensions, see
\cite{Green}). Consider first the anomalies of the
unbroken $SU(3)_C \times U(1)_Q$ part of the gauge group. A
necessary condition for the consistency of the $6$-dimensional
theory is the cancellation of the irreducible gauge anomalies ({\it
i.e.}, which cannot be cancelled by the Green-Schwarz mechanism
\cite{Green:1984sg} or its generalization \cite{Sagnotti:1992qw}
with multiple antisymmetric tensors), required for allowing the
massless gluon and photon. The $U(1)_Q [SU(3)_C]^3$ gauge anomaly
is the only irreducible one,\footnote{The quartic anomaly is
factorizable for $SU(3)$ and $SU(2)$, and irreducible for $SU(n)$
with $n \ge 4$.} and imposing its
cancellation within a generation we find that $\CQ$ should have
opposite chirality compared with $\CU$ and $\CD$.

The $6$-dimensional gravitational and mixed gauge-gravitational
anomalies must also cancel to allow a massless graviton. The
cancellation within one generation of the $6$-dimensional
$U(1)_Q$-gravitational anomaly implies that $\CL$ and $\CE$ also
have opposite chirality. The pure gravitational anomaly cancels
only if the number of fermions with + and $-$ chiralities is the
same (in six dimensions, a selfdual antisymmetric tensor has
gravitational anomaly  equal to that of 28 Weyl fermions, hence
it can not be used to cancel the gravitational
anomaly of $\CL$ and $\CE$). As a result, there must exist an
additional fermion, $\CN$, with the same chirality as $\CE$. The
unprojected zero-mode of $\CN$ may be identified with a
right-handed neutrino. The above arguments yield four possible
chirality assignments of the fermions:
\bear
\CQ_+ \, ,\;\CU_- \, ,\; \CD_- \, ,\; \CL_- \, ,\; \CE_+ \, ,
\; \CN_+ \, ,
\label{first}
 \\
\CQ_+ \, ,\;\CU_- \, ,\; \CD_- \, ,\; \CL_+ \, ,\; \CE_- \,  ,
\; \CN_- \, ,
\label{second}
\eear
and the ones obtained by interchanging + and $-$. With these
assignments, the reducible  anomalies involving $U(1)_Q$ and
$SU(3)_C$ also vanish, because the fermion representations are
vector-like under these groups.

The $SU(2)_W \times U(1)_Y$ 6-dimensional anomalies do not cancel
with the standard model field content, but this may not be
troublesome because the electroweak symmetry is broken. In other
words, one could speculate that the $SU(2)_W \times U(1)_Y$
anomalies from the underlying higher-dimensional theory would be
responsible for  part (or even all!) of the $W$ and $Z$ masses.
Nevertheless, embedding the 6-dimensional theory in a consistent
high-energy theory that includes quantum gravity most likely
requires the $SU(2)_W \times U(1)_Y$ anomalies to be cancelled
within that underlying theory.  This can be achieved  through
the Green-Schwarz mechanism: the $[SU(2)_W]^4, \, [U(1)_Y]^4$,
\, $[SU(2)_W]^2 [SU(3)_C]^2$, $[SU(3)_C]^2
[U(1)_Y]^2$, and $[SU(2)_W]^2 [U(1)_Y]^2$ anomalies\footnote{The
cubic anomaly for $SU(2)$  is identically zero, while the
irreducible $[SU(3)_C]^3 U(1)_Y$ anomaly vanishes within each
generation.}  are cancelled by two antisymmetric tensor
 fields with appropriate Green-Schwarz couplings. We note that
 the presence of reducible anomalies is rather generic in 
 6-dimensional chiral theories that embed the standard model,
 hence antisymmetric tensors are a likely ingredient
 of any realistic 6-dimensional model.\footnote{
 Some components of the antisymmetric tensors survive the
 orbifold projection and have axion-like couplings in the 4-dimensional
 theory. They may acquire masses
 after the compactification, {\it e.g.}, from
terms localized at the orbifold fixed points, or they could provide a
solution to the strong CP problem.}

The main point we make here is that there is an additional
constraint. In six dimensions there are global gauge anomalies,
analogous to the 4-dimensional Witten anomaly \cite{Witten:1982fp}.
They only occur for $SU(3)$ \cite{Elitzur:1984kr}, as well as
$SU(2)$ and $G_2$ gauge theories \cite{Kiritsis:1986mf}; see also
\cite{Bershadsky:1997sb}. Global anomalies are due to the change of
sign of the Weyl fermion determinant under gauge transformations
that are topologically disconnected from the identity; in six
dimensions these arise whenever the gauge group $G$ has nontrivial
$\pi_6(G)$ (the homotopy group of maps of the 6-sphere onto the
gauge group). The mathematical consistency of the theory requires
these to cancel. Since the 6-dimensional $SU(3)_C$ fermion
representations are vector-like, the $SU(3)_C$ global anomaly is
cancelled within each generation. On the other hand, the $SU(2)_W$
global anomaly cancellation condition \cite{Bershadsky:1997sb}
requires
\be
N(2_+) - N(2_-) = 0 \; {\rm mod} \; 6 ~,
\label{z12}
\ee
where $N(2_\pm)$ is the number of doublets of chirality $\pm$.
Since $N(\CQ) = 3$ and $N(\CL) = 1$, the $SU(2)_W$ global anomaly
does {\it not} cancel within one generation for any chirality
assignment.
We are led then to consider the case of $n_g$
generations with identical chirality assignments.
The assignments obtained above, (\ref{first}) and (\ref{second}), give
\be
n_g =  0 \; {\rm mod} \; 3 ~.
\label{ng}
\ee
This is a remarkable result. It is a compelling theoretical
explanation for the existence of three generations.
Although anomaly cancellation in six dimensions allows
the number of generations to be
a multiple of three, the only reasonable prediction is $n_g =3$:
a world with $n_g =0$ would be rather dull, while $n_g \ge 6$
would imply that the gauge couplings blow up very fast above
the compactification scale.

For $n_g = 3$ the effective 6-dimensional
theory is perturbative and well defined for a range of energies
above $1/R$.
The Kaluza-Klein modes of the standard model in $\delta=2$ universal
extra dimensions contribute
at each mass level with $2\times (81/10, \, 11/6, \, -2)$
to the one-loop coefficients of the
$\beta$-functions for the $U(1)_Y$, $SU(2)_W$ and
$SU(3)_C$ gauge couplings. It follows that the 6-dimensional
standard model gauge interactions become non-perturbative at a
scale $\sim 5/R$ \cite{Appelquist:2000nn}. The heavy states of string
theory may become relevant at that scale if the other four extra
dimensions have a large volume
\cite{Arkani-Hamed:1998rs}. Alternatively,
it is conceivable that the 6-dimensional
$SU(3)_C \times SU(2)_W \times U(1)_Y$ gauge couplings approach a
strongly-interacting
fixed point in the ultraviolet \cite{Hashimoto:2000uk},
so that the scale of quantum gravity
need not be lowered much below the Planck scale
by large extra dimensions.

There is also some experimental evidence in favor of our $n_g=3$
prediction: the existence of a fourth generation of chiral fermions
is ruled out at the 97\% confidence level 
(assuming no other physics beyond the standard model) by the
electroweak precision measurements at LEP, SLD and Tevatron
\cite{Groom:2000in}. Moreover, the number of light neutrinos that
couple to the $Z$ was measured at LEP to be very close to three.
However, loopholes in these experimental constraints are not hard
to imagine. For instance, the isospin-violating effects due to the
Kaluza-Klein modes of the top-quark give a positive $T$ parameter
\cite{Appelquist:2000nn}, which in turn may allow a large mass
splitting within the $SU(2)_W$-doublet fermions of a fourth
generation. In this case, a chiral fourth generation would render an
acceptable fit to the electroweak data. Likewise, the constraint on
the number of $SU(2)_W$-charged neutrinos does not apply when they
are heavier than half the $Z$ mass. Hence, the determination of
$n_g$ from anomaly cancellation can be viewed as a prediction that
will be tested in future experiments.

The anomaly cancellation conditions do not restrict the number of
vector-like generations. Even if these exist, there is a simple
reason why they have not been seen yet: their masses are gauge
invariant and are likely to be of the order of the fundamental
(string) scale, $M_s > 1/R$. However, it is also possible to have a
vector-like 6-dimensional generation and 4-dimensional chirality
introduced by the orbifold compactification such that the zero
modes of the 6-dimensional fermions form two chiral generations.
Other ways of cancelling the anomalies can also be found when the
chirality assignments differ between generations. An example is two
generations where one has chirality assignments given by
Eq.~(\ref{first}), while the other's chirality is like that in
Eq.~(\ref{second}). Depending on the 4-dimensional chirality ($L$
or $R$) assigned to the zero-modes by the orbifold, there are two
cases (up to the overall interchanges $+ \leftrightarrow -$ or $L
\leftrightarrow R$):

a) \ \ \parbox[t]{4in}{
$(\CQ_+^1)_L$, $(\CU_-^1)_R$, $(\CD_-^1)_R$, $(\CL_-^1)_L$,
$(\CE_+^1)_R$, $(\CN_+^1)_R$, \\
$(\CQ_+^2)_R$, $(\CU_-^2)_L$, $(\CD_-^2)_L$, $(\CL_+^2)_R$,
$(\CE_-^2)_L$, $(\CN_-^2)_L$ ~,}

\vspace*{2mm}
b) \ \ \parbox[t]{4in}{ $(\CQ_+^1)_L$, $(\CU_-^1)_R$,
$(\CD_-^1)_R$, $(\CL_-^1)_L$, $(\CE_+^1)_R$, $(\CN_+^1)_R$, \\
$(\CQ_+^2)_L$, $(\CU_-^2)_R$, $(\CD_-^2)_R$, $(\CL_+^2)_L$,
$(\CE_-^2)_R$, $(\CN_-^2)_R$ ~,} \\ [3mm] where the upper index
labels the generation. Case a) gives rise only to vector-like
quarks and leptons in the effective 4-dimensional theory. In case
b) however, the zero-modes form two identical generations of chiral
fermions. Thus, a more precise formulation of our result is that
6-dimensional anomaly cancellation requires the existence of more
than one fermion generation, and in the case of {\it identical}
generations ({\it i.e.}, same charges and chiralities, and also
same properties under the orbifold
transformation) their number has to be a multiple of three.

The results obtained so far apply only to non-supersymmetric
theories. In the case of minimal supersymmetry in six dimensions,
the anomalies are significantly more restrictive
\cite{Sagnotti:1992qw}. This is because $(1,0)$ supersymmetry
requires all matter fermions to have the same chirality, opposite
to that of the gauginos  and gravitino. Cancelling the anomalies by
the Green-Schwarz mechanism severely constrains the
 matter content. Thus, for an $SU(3)$ gauge theory with
hypermultiplets only in the $3$ and $\bar{3}$ representations,
the cancellation of local and
global anomalies combined requires that the number of
hypermultiplets be 0, 6, 12, or 18; for $SU(2)$ only 4, 10,
or 16 doublets are allowed.\footnote{The upper limit holds in the
theory without gravity. Using the antisymmetric tensor from the
graviton supermultiplet to cancel the anomaly relaxes the upper
limit and allows for a larger number of hypermultiplets, with the
same periodicity, {\it e.g.}, 24, 30, etc., for $SU(3)$.} Since the
number of fermions in the fundamental representation is $4n_g$ and
$4n_g + 2$, for $SU(3)_C$ and $SU(2)_W$, respectively, this rules
out the 6-dimensional $(1,0)$ supersymmetric ``standard model" with
any $n_g$. Therefore, the supersymmetric models, often considered
in the literature \cite{Dienes:1998vh}, with quarks and
leptons in the bulk of two extra dimensions are anomalous.
One could try the $n_g = 3$ case with two
additional $SU(2)_W$-doublet hypermultiplets. This theory, however,
suffers an irreducible $U(1)_Y [SU(3)_C]^3$ anomaly.
This and the $U(1)_Y$-gravitational anomaly can be cancelled 
simultaneously only if hypermultiplets with exotic 
$U(1)_Y \times SU(3)_C$ charges are added to the theory, 
which is a significant departure from the standard model.
The higher supersymmetries in six dimensions (which
reduce to $N=4$ supersymmetry in four dimensions) do not allow (at
least for now) a prediction regarding the number of
generations: the $(1,1)$ supersymmetric theory  is vector-like,
while the chiral ($2,0$) theory remains rather mysterious. Hence,
the compelling explanation for the existence of three fermion
generations suggests that supersymmetry is broken at the string
scale (or at least above the compactification scales of additional,
smaller universal extra dimensions).

Another issue is whether the number of generations could be
determined based on global anomaly cancellation conditions when
the number of extra dimensions is larger. From the point of view of
string theory only the cases $\delta = 2, 4, 6$ are
interesting.
Given that the $SU(3)_C$ representations are vector-like within a
generation, only $SU(2)_W$ could have a global anomaly.
The relevant homotopy groups are $\pi_8(SU(2)) = Z_2$
for $\delta = 4$, and $\pi_{10}(SU(2))= Z_{15}$ for $\delta = 6$
(see Ref.~\cite{math}).
The generalization to $2 \le \delta \le 12$
of the global anomaly cancellation condition
given in Eq.~(\ref{z12}) is
\be
c_\delta \left[ N(2_+) - N(2_-) \right] = 0 \; {\rm mod} \; n_\delta ~,
\label{zn}
\ee
where $c_\delta$ is an integer, and $n_\delta$ is the number of
homotopy group elements ($n_\delta = 12,\, 2, \, 15$
for $\delta = 2,4,6$.)
Given that $N(2_+) - N(2_-)$ is even within each
generation, there is no constraint on $n_g$ when $\delta = 4$,
while  the global anomaly poses a severe restriction on $n_g$  
when $\delta = 6$.
Only the case $\delta=2$ is both predictive and viable, 
as a consequence of
the fact that the homotopy group $\pi_6(SU(2)) = Z_{12}$
is large and has an even number of elements.

An important point we should stress is that the global anomaly
cancellation condition we have found applies also if the
6-dimensional gauge group is larger, for example, if  $SU(3)_C
\times SU(2)_W \times U(1)_Y$ is
embedded in a gauge group broken by the compactification.
In such a scenario the global anomaly of
$SU(2)_W$ should appear as a local anomaly.

The arguments we have given apply for any size $R$ of the two
extra dimensions as long as there is a range of scales where
an effective 6-dimensional field theory is valid. However,
the usual hierarchy problem suggests that the compactification
scale should be close to the electroweak scale.
One may view the derivation of the number of
fermion generations based on anomaly cancellation conditions
as evidence for the existence of two universal extra dimensions.
Independent support for this conclusion is given by the
successful breaking of the  electroweak symmetry
\cite{Arkani-Hamed:2000hv}
by a composite Higgs field that arises due to
standard model gauge dynamics in two universal extra dimensions.

To summarize, we have shown that global anomaly cancellation
for the standard model in two universal extra dimensions
implies that there must be more than one generation of quarks and
leptons, and if these generations are identical
from the point of view of the fermion
charges, 6-dimensional chiralities, and transformation
properties under the orbifold projection,
then their number should be three.

\bigskip

 {\bf Acknowledgements:} \ We would like to thank Tom Appelquist,
 Hsin-Chia Cheng, Eduardo Pont\'{o}n, and
 Matt Strassler for helpful conversations.
This work  was supported by DOE under contract DE-FG02-92ER-40704.

 \vfil \end{document}